\documentclass[twoside]{article}
\usepackage{fleqn,espcrc2}

\usepackage{graphicx}

\newcommand{\AmS}{{\protect\the\textfont2
  A\kern-.1667em\lower.5ex\hbox{M}\kern-.125emS}}

\title{Systematic Study of Magnetic Interactions in Insulating Cuprates}

\author{Yoshiaki Mizuno, Takami Tohyama and Sadamichi Maekawa\address{Institute for  Materials Research, Tohoku University, Sendai 980-8577, Japan.}
}

\begin{document}

\begin{abstract}
The magnetic interactions in one-dimensional, two-dimensional (2D) and ladder cuprates are evaluated systematically by using small Cu-O clusters.  We find that the superexchange interaction $J$ between nearest neighbor Cu spins strongly depends on Cu-O structure through the Madelung potential, and in 2D and ladder cuprates there is a four-spin interaction $J_{\rm cyc}$, with magnitude of 10 \% of $J$.  We show that $J_{\rm cyc}$ has a strong influence on the magnetic excitation in the high-energy region of 2D cuprates.
\vspace{1pc}
\end{abstract}

\maketitle

A variety of insulating cuprates affords us an opportunity to study the magnetic properties of low-dimensional systems.  Recent experiments for insulating cuprates have revealed interesting characteristics of magnetic interactions: The superexchange interaction between nearest neighbor Cu spins $J$ remarkably depends on Cu-O network structure \cite{Mizuno1}, and additional interactions  such as a four-spin (4S) interaction are important for ladder \cite{Matsuda} and two-dimensional (2D) cuprates \cite{Lorenzana}.  These characteristics indicate the necessity to establish proper magnetic descriptions for the cuprates.  In this paper we perform a systematic study of magnetic interactions for one-dimensional (1D), 2D, and ladder cuprates theoretically.

A starting model to describe the electronic states of cuprates is the $d$-$p$ model, in which hopping integrals between Cu3$d$ and O2$p$ orbitals ($T_{pd}$) and between O2$p$ orbitals ($T_{pp}$), an energy-level separation between the Cu3$d$ and O2$p$ orbitals ($\Delta$), and Coulomb interactions on Cu and O sites are taken into account.  $T_{pd}$ and $T_{pp}$ are obtained by considering not only the bond length dependence but also the effect of the Madelung potential around Cu and O ions.  We find that the potential enhances the magnitudes of $T_{pd}$ and $T_{pp}$ in the 1D cuprates as compared with those in the 2D ones \cite{Mizuno1}.  In the two-leg ladder compounds such as SrCu$_2$O$_3$, $T_{pp}$ along the leg of the ladder is enhanced by the Madelung potential due to adjacent two-leg ladders.  These enhancements play an important role in the dependence of $J$ on the dimensionality.  The $\Delta$ is determined from the difference in the Madelung potential between Cu and O sites.

The magnetic interactions are evaluated by mapping the lowest several eigenstates of small clusters (Cu$_2$O$_7$, Cu$_4$O$_{12}$, and Cu$_6$O$_{17}$) for the $d$-$p$ model onto those of the corresponding Heisenberg-type model \cite{Mizuno3}. For 2D systems, we take into account not only $J$, but also a diagonal interaction $J_{\rm diag}$ and 4S interaction $J_{\rm cyc}$ in the model:
$H= J \sum_{\langle i,j \rangle} {\bf S}_i \cdot {\bf S}_j+ J_{\rm diag}\sum_{\langle \langle i,j \rangle \rangle} {\bf S}_i \cdot {\bf S}_j
+J_{\rm cyc} \sum_{\rm plaquette}(P_{ijkl}+P_{ijkl}^{-1})$,
where ${\bf S}_i$ is a spin operator at site $i$, and $J_{\rm cyc}$ is defined as the coefficient of the 4S cyclic permutation operators $P_{ijkl}$ and $P_{ijkl}^{-1}$, which can be rewritten by using the two-spin interaction $({\bf S}_i \cdot {\bf S}_j)$ and the four-spin interactions $({\bf S}_i \cdot {\bf S}_j)({\bf S}_k \cdot {\bf S}_l)$.  For ladder systems, we distinguish between the nearest neighbor interactions along the leg ($J_{\rm leg}$) and along the rung ($J_{\rm rung}$) of the ladder.

The calculated results are summarized in Table~1, where we take La$_2$CuO$_4$, SrCu$_2$O$_3$, and Sr$_2$CuO$_3$ as typical systems of 2D, ladder and lD cuprates, respectively (see Refs.~1 and 5 for the parameters used in the calculations).
\begin{table}[h]
\caption{The magnetic interactions obtained by mapping the eigenstates of Cu$_2$O$_7$, Cu$_4$O$_{12}$, and Cu$_6$O$_{17}$ clusters for Sr$_2$CuO$_3$, La$_2$CuO$_4$, and SrCu$_2$O$_3$, respectively, onto the corresponding Heisenberg-type models. The numbers in parentheses denote the deviation in the last significant digit.}
\vspace{10pt}
\begin{tabular}{cccc}
\hline
Material & $J$ [eV] & $J_{\rm diag}$ [eV] & $J_{\rm cyc}$ [eV] \\
\hline
Sr$_2$CuO$_3$ & 0.17 & $-$ & $-$ \\
La$_2$CuO$_4$ & 0.146(1) & 0.00(0) & 0.011(1) \\
SrCu$_2$O$_3$ & $J_{\rm leg}$: 0.195(5) & 0.003(2) & 0.018(2)\\
&$J_{\rm rung}$: 0.15(2)&& \\
\hline
\end{tabular}
\end{table}
We find that $J$ in the 1D cuprate is larger than that in the 2D one.  This is caused by the enhancement of the hopping integrals in 1D cuprates as mentioned above.  For 2D cuprates, we obtain $J$ to be $\sim$0.15~eV, consistent with the experimental values \cite{Mizuno1}.  In addition, we find that $J_{\rm cyc}$ is 7 \% of $J$, while $J_{\rm diag}$ is zero.  These results are consistent with a previous cluster calculation \cite{Schmidt}, and a recent analysis of a multimagnon spectrum \cite{Lorenzana}.  For ladder cuprates, we obtain $J_{\rm leg}/J_{\rm rung}$=1.3.  The enhancement of $T_{pp}$ along the leg of the ladder is the origin of the relation $J_{\rm leg}$$>$$J_{\rm rung}$.  $J_{\rm cyc}$ is 10 $\%$ of $J_{\rm leg}$.  Note that the ratio $J_{\rm leg}/J_{\rm rung}$ is smaller than that believed so far, i.e., $J_{\rm leg}$/$J_{\rm rumg}$$\sim$2 \cite{Johnston}.  However, the Heisenberg ladder with the present values of $J_{\rm leg}$, $J_{\rm rung}$, $J_{\rm cyc}$ and $J_{\rm diag}$ reproduces very well the experimental results of the temperature dependence of the magnetic susceptibility (not shown here).  Therefore, we consider the values shown in Table~1 to be reasonable.  Here we would like to emphasize that $J_{\rm cyc}$ plays a crucial role in obtaining the good agreement between the experimental and theoretical magnetic susceptibilities \cite{Mizuno3}.

Next, in order to examine the effect of $J_{\rm cyc}$ on the magnetic excitation, we calculate the dynamical spin-correlation function $S({\bf q},\omega)$ for 2D cuprates.  Figure~1 shows the dispersion and the intensity of $S({\bf q},\omega)$ for a 4$\times$4 Heisenberg model with $J$=0.146~eV and $J_{\rm cyc}$=0.011~eV.  For comparison, the result for $J_{\rm cyc}$=0 are also shown.  We find that the intensity is not sensitive to $J_{\rm cyc}$, while the dispersion is strongly suppressed in the high-energy region.  In particular, it is worth noting that $\omega({\bf q})$ at ${\bf q}$=($\pi$/2,$\pi$/2) becomes smaller than that at ${\bf q}$=($\pi$,0).  This is in contrast with the case of $J_{\rm cyc}$=0, in which the magnetic zone boundary (${\bf q}$=($\pi$,0)$\to$($\pi$/2,$\pi$/2)) has a flat dispersion.  Therefore, it is desirable that inelastic neutron-scattering experiments in the wide energy region be performed to verify the role of $J_{\rm cyc}$, that is, the suppression of the dispersion at ($\pi$/2,$\pi$/2).


\begin{figure}
\includegraphics[width=7.5cm]{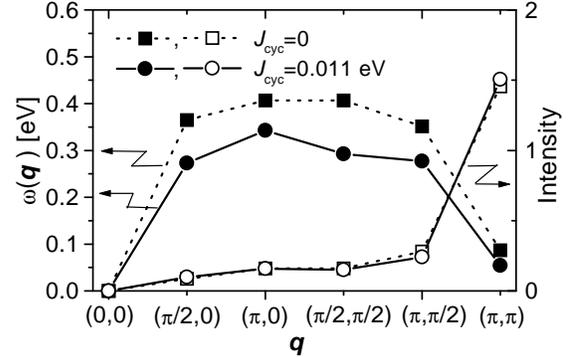}
\vspace{-32pt}
\caption{The magnetic excitation spectra of a 4$\times$4 Heisenberg model with $J$ and $J_{\rm cyc}$.}
\label{fig:1}
\end{figure}

In summary, we have evaluated the magnetic interactions in various cuprates systematically. We have shown that an ionic nature inherent in insulating cuprates is important for the material dependence of $J$. We found that $J_{\rm cyc}$ is $\sim$10 \% of $J$, and greatly influences the magnetic excitation spectra in 2D cuprates.

This work was supported by a Grant-in-Aid for Scientific Research on Priority Areas from the Ministry of Education, Science, Sports and Culture of Japan, CREST and NEDO. The parts of the numerical calculation were performed in the Supercomputer Center in ISSP, University. of Tokyo, and the supercomputing facilities in IMR, Tohoku University.

\end{document}